\documentstyle[preprint,aps,prd]{revtex}
\tighten

\def\bfl{\begin{flushleft}}
\def\efl{\end{flushleft}}
\def\bfr{\begin{flushright}}
\def\efr{\end{flushright}}
\def\bc{\begin{center}}
\def\ec{\end{center}}
\def\be{\begin{equation}}
\def\ee{\end{equation}}
\def\ba{\begin{eqnarray}}
\def\ea{\end{eqnarray}}
\def\baa#1{\begin{array}{#1}}
\def\eaa{\end{array}}
\def\nn{\nonumber }
\def\lb#1{\label{#1}}

\def\text#1{\mbox{#1}}
\def\drm{\text{d}}

\def\Eq{eq. }
\def\Eqs{eqs. }

\def\csii#1#2#3{\Gamma^{#1}_{~ #2 #3} }

\def\eci#1#2#3{K_{#1 #2 #3}}
\def\ecii#1#2#3{K^{#1}_{\ #2 #3} }
\def\eciii#1#2#3{K^{#1 #2}_{\ \ \, #3} }

\def\ececai#1#2#3#4#5#6{\ecii{#1}{#2}{[#3} \ecii{#4}{#5]}{#6}}
\def\ececaii#1#2#3#4#5#6{\ecii{#1}{#2}{[#3} \eciii{#4}{#5}{#6]}}
\def\rmii#1#2#3#4{R^{#1}_{~ #2 #3 #4} }
\def\rmiii#1#2#3#4{R^{#1 #2}_{\,\;\; #3 #4} }

\def\csjiii#1#2#3{[\Gamma]^{#1 #2}_{\ \  #3} }
\def\ecjiii#1#2#3{[K]^{#1 #2}_{\ \  #3} }

\def\sset#1#2#3{S^{(#1)#2}_{\ \ \ #3}}

\begin{document}

%%\wideabs{
%\preprint{}
\draft

\title{\large\bf
(D-E)-dimensional brane worlds and de Rham distribution formalism:
Singular split
versus compactification, 
restrictions on scenario
and revision of gravitational energy problem
}

\author{Konstantin G. Zloshchastiev}

\address{Department of Physics, National University of Singapore,
Singapore 119260, Republic of Singapore\thanks{On leave from {\it 
Department of Theoretical Physics, Dnepropetrovsk State University,
Ukraine}.
Electronic 
address: postmaster@zloshchastiev.itgo.com} 
}

\date{~Received ~~~~~~~~~~~~~~~~~~~~~~~}
%\date{Received \today}
\maketitle

\begin{abstract}
\footnotesize
The proposed so far brane-world cosmological scenarios are
concerned with (D-1)-dimensional embeddings into the D-dimensional spacetime,
besides, it is supposed D=5 as a rule. However, the regarding of the
five-dimensional spacetime as a physical one is a step in past because the
modern concepts of superstring theory require to consider our four-Universe as
a region inside of a much more higher-dimensional manifold. So, it would be
much more realistic to consider our four-Universe as $4$-shell or 3-brane
inside, e.g., 10-dimensional (or even infinite-dimensional) spacetime. 
In turn it immediately means that 
the theory of the $(D-D_E)$-dimensional
singular embeddings, where the number of extra dimensions $D_E > 1$,
is needed. 
Hence, the aim of this work is to provide such a theory: 
we construct the rigorous general
theory of the induced gravity on singular submanifolds. 
At first, we perform
the decomposition of the tangent bundle into the two subbundles which will be
associated later with external and visible (with respect to some
low-dimensional observer) parts of the high-$D$ manifold. 
Then we go to physics
and perform the split of the manifold 
(in addition to the split of the tangent
bundle) to describe both the induced internal geometry and external 
as-a-whole
dynamics of singular embeddings, 
assuming matter being confined on the singular
submanifold but gravity being propagated through the high-$D$ manifold.
With the use of the de Rham axiomatic approach to delta-distributions
we demonstrate that the four-Universe can be singularly embedded 
only in five- and six-dimensional space so if we want to
consider it's embedding 
in 10D then extra dimensions must be included as a product space only.
We
discuss the revealed generic features of the theory such as the multi-normal
anisotropy, restrictions on an ambient space, 
reformulation of the conserved
gravitational stress-energy tensor problem, etc.
\end{abstract}

%%}

%Keyword(s): multiscalar field theory, soliton, brane, quantization, dilaton gravity}

\narrowtext
%\scriptsize
%\footnotesize

\section{Introduction}\lb{s-int}

The theory of pregeometry, i. e., induced rather than imposed
gravity, proposed in late 60's by Sakharov \cite{sak}
(see also the dedicated review \cite{amve})
is revived nowadays in the another context - in the higher-dimensional
models of the Universe.
At early stages of this mainstream the Sakharov's
idea was applied to n-dimensional manifolds
\cite{ndig}, as a rule
in the connection with the Kaluza-Klein paradigm (i.e., assuming
the further compactification of extra dimensions).
At the same time there appeared the 
``Universe as vortex'' model \cite{akam} 
which demonstrated that the KK compactification is not
the only way of the ``hiding'' of extra dimensions but that idea
remained almost unnoticed on the background of the universal
popularity of KK-type theories.

Much later, 
being inspired by the two works of Rubakov and Shaposhnikov \cite{rush} 
(the first paper was devoted
to the ``Universe as domain wall'' conjecture whereas the second one
proposed a high-dimensional solution of the hierarchy problem), 
Gogberashvili \cite{gog} and later independently of him
Randall and Sundrum \cite{rasu} have
put forward the $(4+1)$-dimensional
brane-world cosmological scenarios with the emphasis
on the hierarchy problem.
Despite the difference of approaches (Gogberashvili used
the geometrical junction theory \cite{sen,dau,isr,mtw}
whereas Randall and Sundrum did the variational method with the
stress-energy tensor containing delta-functions),
terminology (Gogberashvili called it 
as ``Universe as shell'' scenario
whereas Randall and Sundrum used the modern 
``Universe as three-brane'') and physical assumptions,
all three authors, in fact, proposed the same point of view
which amplified the interest \cite{aadd0} to the high-dimensional
cosmological models where extra dimensions were assumed to be 
orthogonal to the Universe as a singular shell or $3$-brane rather
than compactified.

The further research efforts were directed toward the 
diversifying of the
physically specific models of five-dimensional brane-world cosmology
as well as toward the elucidation of
the relations between the two above-mentioned 
approaches \cite{cpr,sms,dedo}
(the class of branes belongs to much more wide family of singular
shells; the integration of Einstein equations with the distributional 
sources can be reformulated in terms of the more rigorous
junction formalism \cite{makh}, etc.), 
the generalizing in several aspects \cite{dedo}, and the
considering of high-dimensional brane models, e.g., $D8$-branes 
(Dirichlet $9$-embeddings) in a
$10$-dimensional SUGRA spacetime \cite{cpr}.

Overlooking the geometrical achievements of the ``brane-world rush''
one can reveal that all the studies are concerned with
$(D-1)$-dimensional embeddings and their special case, $(D-2)$-branes,
into the $D$-dimensional spacetime, besides it is supposed $D=5$ as a rule.
However, the serious considering of five-dimensional spacetime
means, in fact, step back in past because the modern concepts of
superstring theory require to consider our four-Universe
as a region inside of a much more high dimensional spacetime.
In turn, it immediately means that we are needed in the theory
of the geometrically induced gravity on 
$(D-D_E)$-dimensional embedded (singular) manifolds where
the number of extra dimensions $D_E > 1$.
To the best of our knowledge such a singular junction 
formalism has not yet been presented
anywhere despite the embedding of Riemann surfaces into 
a higher-dimensional spacetime is a well-studied classical 
problem \cite{embe}.
The aim of this work is thus to equip a reader with it.

The emphasis will be done on the fundamental aspects of the
theory because the early falling into physical particularities can 
raise some confusing whether discussed properties are generic or not.
Thus, we will try to construct the (more or less)
rigorous general formalism of the singular submanifold theory
founding on the  geometrical junction theory
because  the 
straightforward integration of Einstein equations with distributions 
is good for quick obtaining of certain results rather than
for full understanding of what we are doing.
Throughout the paper we will emphasize on the difference between
geometries of the Kaluza-Klein type models and the theory of
singular submanifolds.

In Sec. \ref{s-dec} we perform the decomposition of the tangent
bundle into the two subbundles which will be associated later
with external and visible (with respect to some low-dimensional
observer) parts of the high-$D$ manifold.
This section is common for both the KK models and the 
singular submanifold
theory, and provides us with underlying mathematical language.
In Sec. \ref{s-ssm} we go to physics and
perform the split of the manifold (in addition
to the split of the tangent bundle) to
describe the induced internal  geometry and external as-a-whole
dynamics of singular embeddings.
We assume matter (including the Standard Model with its fiber bundles)
to be confined on the singular (sub)manifold and introduce the
multi-normal surface stress-energy tensor.
We consider both the direct (naive) approach and the axiomatic
theory of delta-like distributions based on the de Rham currents.
In Sec. \ref{s-dis} we discuss the features of the 
brane world viewpoint in comparison 
with those of the KK models.

\section{($V+E$)-decomposition of tangent bundle}\lb{s-dec}

Let us consider the $D$-dimensional Riemann manifold $\Sigma$ 
assuming
$T_\Sigma$ is its underlying tangent bundle space.
Let us cover the bundle by $D$ basis vectors ${\bf e}_\alpha$
(here and below Greek indices run from $1$ to $D$).
Further, let us divide the set 
$\{ {\bf e}_\alpha,~\alpha= 1,~2, ..., D \}$ into 
the two subsets 
$\{ {\bf e}_i, ~ i= 1,~2, ..., D_V \}$ and
$\{ {\bf e}_a, ~ a= D_V+1, ..., D_V+D_E \}$ where
$D_E + D_V = D$.
Further it will be everywhere assumed  
that Latin indices $i,j,k,l,m$ run from $1$ to $D_V$,
$a,b,c,d,f$ do from $D_V+1$ to $D_V+D_E=D$.

Then, if there are imposed some mathematical rules for the dividing the
set  $\{ {\bf e}_\alpha \}$ 
into the sum $\{ {\bf e}_i \} \cup \{ {\bf e}_a \}$ 
it means
that $T_\Sigma$ is decomposed into the two subbundles which we will
call as $T_{E(xtra)}$ and $T_{V(isible)}$ keeping in mind the
forthcoming physics which will be based on this formalism.
In reality, it is enough to restrict ourselves by the case $D_V=4$
but in this paper we will study the most general 
case of arbitrary $D_E$ and $D_V$.

Note, the decomposition $T_\Sigma = T_{E} \oplus T_{V}$
does not mean yet the split of the $D$-dimensional manifold $\Sigma$ 
into the sum of
the (singular) submanifolds $E$ and $V$.
At this stage we just have ($V+E$)-relabeled the underlying bundle space 
of $\Sigma$ to obtain some useful basic formulae which in their turn will
gain a concrete physical sense {\it only}
when, running ahead, considering the related physical entities, singular submanifolds.                   

The ($V+E$)-decomposition of tangent bundle space is the natural
generalization of the ($V+1$)-decomposition, the basic formalism of
($D-1$)-embeddings in $D$-spacetime, on the case $D_E > 1$.
The ($V+1$)-decomposition (especially its special cases $3+1$ and $4+1$)
happened to be excellent language for singular shell theory, Hamiltonian
formulation of general relativity and
Cauchy problem in GR, but, as was mentioned above,
the modern concepts demand for the 
language for the description of more ``compact'' ($D_V < D-1$) embeddings 
into high-$D$ spacetime.

Further, for simplicity we will suppose the basis 
$\{ {\bf e}_\alpha  \} = \{ \{ {\bf e}_i \}, \{ {\bf e}_a \} \}$
to be orthogonal and commutative,
\be
{\bf e}_a \cdot {\bf e}^b = \delta_a^b,~
{\bf e}_i \cdot {\bf e}^k = \delta_i^k,
                                                                 \lb{eq01} 
\ee
besides we will assume the block-orthogonality condition
\be
{\bf e}_a \cdot {\bf e}_i = 0.
\ee
Assuming that the connection is symmetric and compatible with metric
we can decompose it into $V(isible)$ and $E(xtra)$ parts as well:
\be
\nabla_{{\bf e}_\nu} {\bf e}_\mu  = 
\ecii a \nu \mu {\bf e}_a  +
\ecii i \nu \mu {\bf e}_i,~~
\eci \alpha \nu \mu \equiv 
{\bf e}_\alpha \cdot \nabla_{{\bf e}_\nu} {\bf e}_\mu,
\ee
and one can check that connections 
with non-mixed ``$E, V$''-indices coincide
with the Christoffel symbols in the corresponding subbundle $V$ or $E$:
\be
\ecii a b c = ~^{(E)}\!\csii a b c , ~~~
\ecii i j k = ~^{(V)}\!\csii i j k ,
\ee
so that below when dealing with Christoffel symbols we will omit 
the superscripts $^{(V)}$ and $^{(E)}$ for brevity.
With this in hands we can $(V+E)$-decompose all the necessary tensors.
For some needed components 
of the Riemann tensor in natural frame we hence have
\ba
&&\rmii i j k l = ~^{(V)}\!\rmii i j k l + \ececai i a k a l j ,\\
&&\rmii a j k l = 
\ecii{a}{j}{[l ; k]} + \ececai a \lambda l \lambda k j.    \lb{eq06}
\ea 
These expressions 
are the generalizations of the Gauss-Codacci equations of the
$(V+1)$-decomposition which in turn is the underlying formalism both
for the singular shell theory in the ordinary spacetime 
$D=4$ \cite{mtw} and for the
proposed brane-world (toy) models of the four-Universe as a $3$-brane
in the higher-dimensional space with $D=5$ \cite{sms,dedo}.
Indeed, if the $E$-index $a$ 
has only one value, $a_{D_V + 1} = n$, we obtain
\ba
&&~^{(D)}\!\rmii m i j k = 
~^{(D-1)}\!\rmii m i j k + g^{nn} K_{i [ j} K^m_{k ]},  \\
&&~^{(D)}\!\rmii n i j k = g^{nn} K_{i [k ; j]},                  \lb{eq08}
\ea
where the extrinsic curvature $K_{i j}\equiv\eci{n}{i}{j}$, and 
certain features of the gaussian/synchronous 
reference frame 
were taken into account.
Note, that the equation (\ref{eq06}) 
in comparison with \Eq (\ref{eq08}) contains
the extra (second) term which is caused by the fact $D_E > 1$.
Running ahead, we say that appearance of such terms is inevitable and
sufficiently complicates matter.

The components of the decomposed Einstein tensor are
\ba
&& G^i_k = ~^{(V)}\!G^i_k + \rmiii a i a k - g^i_k \rmiii a j a j
           - \frac{1}{2} g^i_k 
           \left( 
                  ^{(E)}\!R  + 
                  \ececaii l a j a j l + \ececaii a j c j c a
           \right)
           + \ececaii j a k a i j    \, , \lb{eq09}\\
&& G^i_d = R^i_d 
         = \eciii{a}{i}{[d, a]}  + \eciii{j}{i}{[d, j]} +
           \ececaii i \lambda a a \lambda d  +
           \ececaii i \lambda j j \lambda d  \, , \lb{eq10}\\
&& G^c_d =
           ~^{(E)}\!G^c_d + \rmiii c j d j - g^c_d \rmiii a j a j
           - \frac{1}{2} g^c_d 
           \left( 
                  ^{(V)}\!R  + 
                  \ececaii a j b j b a + \ececaii l a j a j l
           \right)
           + \ececaii a j d j c a \, ,                            \lb{eq11}
\ea
where we defined
\[
\rmiii a i b k = 
\eciii a i {[k, b]} + \ececaii i \lambda b a \lambda k  \, .
\]
Now we have all the necessary formulae to consider the geometry and
physics of
$D_V$-dimensional singular embeddings.

\section{Singular split: geometry encounters physics}\lb{s-ssm}

So far the $(V+E)$-decomposition formulae just represented
the split of the tangent bundle hence were nothing but 
the simple relabeling of
the base manifold $\Sigma$.
Now let us suppose that there exists some entity that performs not
only the split of the bundle
$T_\Sigma = T_{E} \oplus T_{V}$ but also the split of the base
manifold into the parts 
$\Sigma= E(xtra) \cup V(isible)$.\footnote{Throughout
the paper 
we will call the former as the
{\it decomposition} or {\it simple split} and the latter as 
the {\it singular split}.}
Each of these two singular submanifolds can be assumed to have 
its own geometry 
and matter on its $D_E$- or $D_V$-dimensional worldsheet.
Besides the intrinsic geometry a singular submanifold can move
as a whole inside the parent 
spacetime $\Sigma$ hence it has own non-trivial external dynamics.
Unlike this, by definition the non-singular manifold has neither
(hyper)surface matter nor external dynamics, and represents itself
just some (relative) region of $\Sigma$ having no physical carrier.

For definiteness, we select for further studying the singular submanifold
$V$ assuming $E$ as the rest, $E=\Sigma/V$.
The embedding $V$ will be associated with our visible 
four-dimensional Universe hence $D_V=4$ but for generality we will assume
arbitrary $D_V < D$.

Thus, the  singular submanifold $V$ appears to be the physical
carrier that ``fixes'' the $(V+E)$-decomposition.
The question now is how to define the intrinsic stress-energy tensor
of the matter on its (hyper)surface.
For the ($D-1$)-dimensional singular embeddings (the standard thin-shell
formalism) we had the 
following definition
of surface stress-energy tensor
\be
~^{(V+1)}\!S^\alpha_\beta = \lim\limits_{\varepsilon \to 0}
                 \int\limits_{-\varepsilon}^\varepsilon
                 T^\alpha_\beta \drm n,                     \lb{eq12}
\ee
where $T^\alpha_\beta$ is the general $D$-dimensional stress-energy
tensor, $n$ is the (only) extra coordinate ``piercing'' the submanifold.
However, now our submanifold $V$ has $D_E$ normals 
towards the $E$(xtra) directions
$\{ {\bf n}^{  a} \} \ (  a = D_V + 1,...,\,  D)$ so the question
now is how to generalize the standard thin-shell concepts.

\subsection{Direct approach reveals 
contradictions}\lb{s-ssm-nai}

The most natural generalization of the integral in \Eq (\ref{eq12})
seems to be the integral 
$\int T^\alpha_\beta \drm n^a$ but it immediately does mean the
appearance of an extra index at $S^\alpha_\beta$ (this is  required
also by the left-hand sides of \Eqs (\ref{eq18}) - (\ref{eq20})).
Then the genuine surface stress-energy tensor is given
by the following sum
\be
S^\alpha_\beta =  \prod\limits_{a=D_V+1}^{D=D_V+D_E} \sset a 
\alpha \beta, \ \ \
\sset a \alpha \beta = \lim\limits_{\varepsilon \to 0}
                 \int\limits_{-\varepsilon}^\varepsilon
                 T^\alpha_\beta \, n^a_\mu \drm x^\mu,
                                               \lb{eq13}
\ee
where ${\bf n}^a$'s are $D-D_V=D_E$ normal vectors to $V$.
From the viewpoint of the observer living inside $V$ this index seems
to be numbering
internal (non-spacetime) degrees of freedom because its 
lowering/raising is governed
by the $E$-metric $g_{a b}$ only, by virtue of the block-orthogonality
condition.
In fact, \Eq (\ref{eq13}) reflects itself the new mechanism of 
generation of internal degrees
of freedom which will be referred throughout the paper as the
{\it multi-normal anisotropy}
and discussed in more details later.

Assuming that the Einstein equations in $D$-spacetime $\Sigma$,
\be
G^\mu_\nu = k_D^2 T^\mu_\nu,                             \lb{eq14}
\ee
are valid, we obtain from \Eq (\ref{eq09}) - (\ref{eq11})
both the equation for induced gravity on $V$,
\be
~^{(V)}\!G^i_k = k_D^2 T^i_k - \rmiii a i a k + g^i_k \rmiii a j a j
           + \frac{1}{2} g^i_k 
           \left( 
                  ^{(E)}\!R  + 
                  \ececaii l a j a j l + \ececaii a j c j c a
           \right)
           - \ececaii j a k a i j    \, ,                         \lb{eq15}
\ee
where all $R$'s are supposed to be the functions of the $D$-dimensional
stress-energy tensor,
and the additional equations
\ba
&&     \eciii{a}{i}{[d, a]}  + \eciii{j}{i}{[d, j]} +
       \ececaii i \lambda a a \lambda d  +
       \ececaii i \lambda j j \lambda d = k_D^2 T^i_d \, , \lb{eq16}\\
&&     ~^{(E)}\!G^c_d + \rmiii c j d j - g^c_d \rmiii a j a j
           - \frac{1}{2} g^c_d 
           \left( 
                  ^{(V)}\!R  + 
                  \ececaii a j b j b a + \ececaii l a j a j l
           \right)
           + \ececaii a j d j c a = k_D^2 T^c_d\,  ,   
                         \lb{eq17}
\ea
which will be also important below.

Further, applying the pill-box integration over all $E$-coordinates
(with successive taking of the limit $\varepsilon \to 0$)
to the
Einstein equations (\ref{eq14}) 
in the decomposed form (\ref{eq15})-(\ref{eq17}) 
and assuming that $\Sigma$-metric is
continuous across $V$  
we obtain that the first derivatives have a finite jump across $V$,
and 
the comparison of the integrands at $\int d^{E-1} x^a$ (which is,
by definition, the integration over all $E$-coordinates except $a$th)
yields the junction conditions 
\ba
&& \ecjiii a i k - \delta^i_k \ecjiii a j j  
   - \frac{1}{2} \delta^i_k \csjiii{[a}{b]}{\ \ b}
   = \frac{k_D^2}{D_E} \sset a i k \, ,                           \lb{eq18}\\
&& \ecjiii a i d 
   - \delta^a_d 
         \ecjiii b i b 
   - \delta^a_d 
   \csjiii j i j
   = \frac{k_D^2}{D_E} \sset a i d  \,   ,                        \lb{eq19}\\
&& \csjiii a c d  
   - \delta^a_d \csjiii b c b
   - \frac{1}{2} \delta^a_d \csjiii{[a}{b]}{\ b} 
   +  \delta^{[ a}_d \ecjiii{c]}{i}{i} 
   = \frac{k_D^2}{D_E} \sset a c d \, ,                            \lb{eq20}
\ea
where it was used that only the derivatives with respect to $x^a$
survive, and the jump ``$[\,]$'' is defined as
\[
[Z] = Z(x^i,n^a_\mu x^\mu=n^a_\mu x^\mu_0 + 0) - 
Z(x^i, n^a_\mu x^\mu=n^a_\mu x^\mu_0 - 0),
\]
where the coordinates $x^{\mu}_0$ point out the position of $V$.
Considering \Eq (\ref{eq13}), the $\Sigma$-stress-energy tensor can
be imagined in the split form as the superposition of 
the $B$(ulk) part and the $V$-surface's part as the
sum of $S$'s over all $D_E$ normals
\be
T^\mu_\nu = B^\mu_\nu +
            \prod\limits_{a}^{} \sset a \mu \nu \, 
\delta (n^a_\mu x^\mu - n^a_\mu x^\mu_0) ,
                                      \lb{eq21}
\ee
where the second term is zero everywhere except $V$.

The system of \Eqs (\ref{eq14}) - (\ref{eq21}) completely determines
both the intrinsic geometry of $V$ and the external dynamics of $V$ 
as a whole inside the parent manifold.
In some cases it can be greatly simplified, e.g., if one assumes
$E$-coordinates to be flat
(i.e., gaussian/synchronous: $g_{ab}=\text{const}$ hence
$\csii a b c = 0$, $\eci a i b =0$, 
$\eci a i j = - (1/2) g_{i j, a}$ , etc.),
and/or if $D$-dimensional manifold admits $Z_2$-symmetry $[Z]= \pm 2 Z$.
 
Everything looks fine but in practice it is not so.
The problem is that the stress-energy tensor given by
\Eqs (\ref{eq13}) and (\ref{eq21}) does not describe the desired 
$V$-embedding.
It describes instead the aggregate of the orthogonal
singular hyperplanes crossing the $V$ submanifold.
Moreover, it seems impossible to reconcile the definition of stress-energy
tensor based on the pill-box integration  with the circumstance
that the L.H.S. (hence R.H.S.) of \Eqs (\ref{eq21}) acquire an
extra $E$-index $a$ if $D_E > 1$.
The reasons of this contradiction 
happen to be quite deep - we have to consider the 
axiomatic theory of delta-like distributions to clear them up.

\subsection{Axiomatic
approach to singular embeddings: de Rham currents}\lb{s-ssm-axi}

Let us start from the very initial definitions - the definitions 
of delta-like distributions within the frameworks of the geometrical theory
of distributions \cite{cho77} 
(we will use also Ref. \cite{stelea}
where that theory was adapted to thin-shell formalism).

Let $C^\infty_0 (R^n)$ be the vector space of compactly supported smooth 
functions on $R^n$. 
If $(x^1,..., x^n)$ are the canonical coordinates on $R^n$, 
we define the operators $D_i=\partial/\partial x^i$,
$D^\alpha = D^{\alpha_1} ... D^{\alpha_n}$, with
$\alpha=\{ \alpha_1, ..., \alpha_n \}$ to be integer non-negative numbers.
The $C^p$ - topology is defined on $C^\infty_0 (R^n)$ by saying that the 
sequence $\varphi_n$ tends to zero if:
there is a compact set $K$ with $\text{supp}(\varphi_n) \subset K$
and $D^\alpha \varphi_n$ vanishes uniformly for $x \in K$ and all $\alpha$ 
satisfying $|\alpha| = \alpha_1 + ... + \alpha_n \leq p$.

Then the {\it distribution} on $R^n$ is the linear map $T$: 
$C^\infty_0 (R^n) \mapsto R$, and it is continuous in the 
$C^\infty$ topology. 
The vector space of distributions on $R^n$ is denoted $D (R^n)$.
Now we extend the operators $D_i$ to the space of distributions by setting 
\[
(D_i T, \varphi) = (- 1)^{|\alpha|} (T, D_i \varphi), \
\forall \varphi \in C^\infty_0 (R^n),
\]
and denote by $\Omega^\ast (U)$ the algebra of exterior forms on $U$ 
where U is a domain of $R^n$. 
Let  $\Omega^q_c (R^n)$ be the space of $q$-forms on $R^n$ with 
compact support 
recalling that the $q$-form with compact support is
$\omega = \sum \omega_{i_1 ... i_q} \drm x^{i_1} ... \drm x^{i_q}$ where
$\omega_{i_1 ... i_q} \in C^\infty_0 (R^n)$, 
$1\leq i_1 \leq ... \leq i_q \leq n$.     

The topological dual of $\Omega^{n-q}{c} (R^n)$ is the space of the
(de Rham)
currents of degree $q$ and is denoted by $D^q (R^n)$, then
the current $T \in D^q (R^n)$ may be considered as a differential form
$T = \sum \omega_I \drm x^I$, $|I|=q$, where $I = {i_1, ..., i_q}$
and distribution coefficients $\omega_I$ are defined by
$(\omega_I, \varphi) = \pm (T, \varphi \drm x^{I^0}) $, 
$\forall \varphi \in C^\infty_0 (M)$, where
$I^0$ is the compound index defined by $\star \drm x^I = \pm \drm x^{I^0}$.
The exterior derivative on smooth forms induces an exterior 
derivative operator on the spaces of currents:
\[
\drm: D^q (R^n) \mapsto D^{q+1} (R^n), \ \
(\drm T, \varphi) = (-1)^q (T, \drm \varphi). 
\]

To see explicitly the links between this formalism and usual definition
of delta-like distributions let us localize the definitions of the de Rham
currents. 
Let $M$ be a smooth manifold of dimension $n$.
If $U$ is an open set in $M$ the space of distributions with support in 
$U$, $D(U)$, is the topological dual of the space $C^\infty_0 (U)$.  
Let $\Omega^\ast_0 (M)$ denote the algebra of compactly supported 
$q$-forms on $M$ and $\Omega^\ast_0 (U)$ do the space of smooth compactly 
supported $q$-forms on $U$. 
Taking the topological duals of these spaces we obtain the space of 
distributions $D (M)$ and the spaces of currents $D^q (M)$ and $D^q (U)$. 
For now on we will consider only space-time manifolds $M$.
Let $S$ be the hypersurface in $M$ defined by the equation
$\eta (x^\alpha) = 0$, $\nabla \eta \not= 0$, where $x^\alpha$
are local coordinates on $M$.
Denote by $\theta (\eta)$ the characteristic function of the 
$n$-dimensional domain $\Gamma = \{ \eta >0 \}$ with the boundary 
$\Sigma = \partial \Gamma$
\[
\theta (\eta) = 
\left\{
\begin{array}{l}
1 \ \ \eta \geq 0 \\
0 \ \ \eta < 0 
\end{array}
\right. ,
\]
so $\theta (\eta)$ defines a current by 
\[
(T_\theta, \varphi) = \int \theta \wedge \varphi =
\int\limits_{\eta \geq 0} \varphi ,
\]
hence
\[
(T_{\drm\theta}, \varphi) = \int \drm \theta \wedge \varphi =
-\int\limits_{\Gamma} \drm \varphi ,
\]
and one can prove that we can uniquely associate with 
$\Sigma$ a closed 1-form $\delta (\Sigma) \in H^1 (M)$ 
(with $H^q (M)$ to be the q-th de Rham cohomology of $M$) so that
$\drm \theta (\eta) = \delta (\eta) \drm\eta$ which denotes that
$D_i \theta(\eta) = D_i \eta \delta (\eta) $.
Thus, the delta ``function'' $\delta (\Sigma)$ has been
defined as the special exact 1-form. 

Finally, let us give the integral representation of the distribution 
$\delta (\Sigma)$. 
Let $\chi (x) \in C^\infty_0 (M)$ be the nonnegative function on $M$,
supported in the vicinity of $\Sigma$ with
$\int \chi (x) d x = 1$.
We set $\chi_\varepsilon (x) = \varepsilon^{-n} \chi (x/\varepsilon)$.
Now if $\text{supp}\chi = K$ then 
$\text{supp}\chi_\varepsilon = K_\varepsilon$ and
$\int \chi_\varepsilon (x) d x = 1$.
Then $T_{\chi_\varepsilon} \to \delta$ as 
$\varepsilon\to 0$ in the sense that 
\[
\lim\limits_{\varepsilon\to 0} (T_{\chi_\varepsilon}, \varphi)
= (\delta, \varphi), \ \forall \varphi\in C^\infty_0 (M), 
\]
i.e. we can approximate the delta-singularities by smooth functions
in this way.
Now we have all the necessary definitions to consider the thin-shell
(singular embedding) theory within the frameworks of de Rham currents' 
approach.\\

{\it Standard ($D_E=1$) thin-shell theory and de Rham currents}\\
Again, let $M$ be a space-time manifold and  $M_+$ and $M_-$ be two 
overlapping domains of this manifold, 
$\Sigma$ be the hypersurface contained in $M_+ \cap M_-$ which 
embedding is defined in local coordinates
by the equation  $\eta (x^\alpha) =0$.
For the pair of metrics $g_{ab}^\pm$ defined on $M^\pm$ respectively,
assuming coordinates be continuous across $\Sigma$, 
we will impose the condition
\be                                        \lb{eContg}
[g_{\alpha\beta}] \equiv g_{ab}^+ - g_{ab}^- \ |_\Sigma =0.
\ee
Also we have
\[
g^{\alpha\beta}_\pm\, \partial_\alpha \eta \partial_\beta \eta=
\varepsilon_\pm (\eta) \alpha_\pm^2 (x),
\]
so 
\ba
&&n_\alpha^\pm = \frac{1}{\alpha_\pm} \partial_\alpha \eta,   \nn\\
&&g^{\alpha\beta}_\pm n_\alpha^\pm n_\beta^\pm |_\Sigma =
\varepsilon_\pm (x) = 
\left\{
       \begin{array}{l}
       0, \ \Sigma \ \text{is lightlike} \\
       \pm 1, \ \text{otherwise}
       \end{array}
\right. .  \nn
\ea
To describe uniformly both null and non-null surfaces let us introduce
the vector $N_\alpha$ such that
\[
N^\alpha n_\alpha^\pm = 1/ \zeta_\pm, \ \ \text{i.e.,} \ \
N^\alpha \partial_\alpha \eta = \alpha/ \zeta_\pm, 
\]
where $\zeta$'s are some functions, for definiteness $\zeta$ is $1$ if
$\Sigma$ is timelike and $-1$ otherwise.
This allows us to consider the case of the null surface $\Sigma$
as the limit $\varepsilon \to 0$ of the non-null one.
In terms of the vielbein $\omega^\alpha$ the metric in $M$ is 
\[
d s^2 = \eta_{\alpha\beta} \omega^\alpha \omega^\beta,
\]
and in presence of $\Sigma$ we have 
\be                                                            \lb{eSplitOm}
\omega^\alpha = \omega^\alpha_+ \theta (\eta) + 
\omega^\alpha_- \theta (-\eta),
\ee
where $\theta$ is the 0-form 
current defined above as the (analogue of the) Heaviside function.
The condition (\ref{eContg}) implies
\[
(\omega^\alpha_+ -
\omega^\alpha_-) |_\Sigma =0,
\]
hence we have 
$\drm \omega^\alpha = \drm \omega^\alpha_+ \theta (\eta) +
\drm \omega^\alpha_- \theta (-\eta)$, and the connection 1-form
is given by the first Cartan equation 
$\drm \omega^\alpha = \omega^\beta \wedge \Gamma^\alpha_\beta$ as
\be                                                 \lb{eGamDecomp}
\Gamma^\alpha_\beta = \Gamma^{\alpha  +}_\beta \theta (\eta) +
\Gamma^{\alpha  -}_\beta \theta (-\eta).
\ee
The second Cartan equation yields the curvature 2-form decomposed
into two bulk ($\pm$) parts and one singular on $\Sigma$
\be                                    \lb{eCurvDecomp}
R^\alpha_\beta = \drm \Gamma^\alpha_\beta + \Gamma^\alpha_\gamma \wedge
\Gamma^\gamma_\beta = R^{\alpha  +}_\beta \theta (\eta) +
R^{\alpha  -}_\beta \theta (-\eta) +
\delta (\eta) S^\alpha_\beta,
\ee
where we have defined
\[
S^\alpha_\beta = 
\drm \eta \wedge [\Gamma^\alpha_{\beta\gamma}] \omega^\gamma,
\]
and
$[\Gamma^\alpha_{\beta\gamma}] \omega^\gamma \equiv 
(\Gamma^{\alpha +}_\beta - \Gamma^{\alpha -}_\beta)|_\Sigma$, and
it was used the definition $\drm \theta = \delta \drm \eta$ from the 
introductory part above.
Further, using $\drm \eta = \alpha n_\gamma \omega^\gamma$
we obtain
\[
S^\alpha_{\beta\gamma\sigma} \omega^\gamma \wedge \omega^\sigma =
- 2 \alpha [\Gamma^\alpha_{\beta\gamma}] n_\sigma 
\omega^\gamma \wedge \omega^\sigma,
\]
or, simply,
\be
S^\alpha_{\beta\gamma\sigma}  =
- 2 \alpha [\Gamma^\alpha_{\beta\gamma}] n_\sigma.
\ee
Let us imply now the coordinate basis.
Then the jump of the first derivative of metric across $\Sigma$
is
\[
[\partial_\mu g_{\alpha\beta}]= \zeta \gamma_{\alpha\beta} n_\mu,
\]
where $\gamma_{\alpha\beta}$ is the jump in the transversal
derivative 
$\gamma_{\alpha\beta} = \alpha N^\mu [\partial_\mu g_{\alpha\beta}] $,
so that the jump of the Christoffel symbols across $\Sigma$ can be expressed
as
\be
[\Gamma^\alpha_{\beta\gamma}] =
\zeta (\gamma^\alpha_\beta n_\sigma +
\gamma^\alpha_\sigma n_\beta - \gamma_{\beta\sigma} n_\alpha)/2,
\ee
so the surface Riemann and Ricci tensors are, respectively,
\ba
&&S^\alpha_{\beta\gamma\delta} = \frac{\alpha}{2} 
\zeta 
\left[
      n^\alpha (\gamma_{\beta\delta}  n_\gamma -
                \gamma_{\beta\gamma}  n_\delta) -
      n_\beta (\gamma^\alpha_\gamma  n_\delta  - 
               \gamma^\alpha_\delta  n_\gamma)
\right],                                            \\
&&S_{\alpha\beta} = \frac{\alpha}{2} 
\zeta 
\left[
      \gamma_\alpha n_\beta + \gamma_\beta n_\alpha
      - \gamma n_\alpha n_\beta 
      - \breve\gamma h_{\alpha\beta} 
      - \varepsilon (\gamma_{\alpha\beta}-\gamma h_{\alpha\beta})
\right],                                            
\ea
where $\gamma^\alpha=\gamma^\alpha_\beta n^\beta$,
$\breve \gamma = \gamma^\alpha n_\alpha$,
$\gamma = \gamma_{\alpha\beta} h^{\alpha\beta}$,
$h_{\alpha\beta}$ is the metric on $\Sigma$.
As was promised we are able to obtain the surface Ricci tensor
for light-like shells by taking the $\varepsilon\to 0$ limit:
\ba
S_{\alpha\beta} = \frac{\eta}{8\pi} 
\left(
      \gamma_\alpha n_\beta + \gamma_\beta n_\alpha
      - \gamma n_\alpha n_\beta 
      - \breve\gamma h_{\alpha\beta} 
\right),                                            
\ea
i.e., the results of ref. \cite{mtw} were completely reproduced.\\

{\it Embeddings with $D_E > 1$ in light of de Rham approach}\\
So far we assumed $D_E=1$, i.e., (D-1)-dimensional layer in D-dimensional
manifold.
Let us turn now to the case of $\Sigma$ with $D_E > 1$
which embedding is described by $D_E$ equations
\[
\eta^{(a)} (x) =0, \ a=1, .., D_E .
\]
The most unexpected thing which appears is that $D_E$ 
{\it cannot be more
than two}!
Indeed, within frameworks of the de Rham approach the N-dimensional
delta-singularity must be described by the N-form
$\drm \theta (\eta^{(1)}) \wedge \drm \theta (\eta^{(2)}) \wedge ..
\wedge \drm \theta (\eta^{(N)}) = 
 \delta (\eta^{(1)}) .. \delta (\eta^{(N)}) \, \drm\eta^{(1)}\wedge ..
\wedge  \drm\eta^{(N)}$.
This must appear in the singular part of the
curvature 2-form (\ref{eCurvDecomp}).
But the latter is a 2-form therefore singular part must be a 2-form
as well, and one cannot insert there the 3-form 
$\delta (\eta^{(1)}) \delta (\eta^{(2)}) \delta (\eta^{(3)}) \, 
\drm\eta^{(1)}\wedge \drm\eta^{(2)}
\wedge  \drm\eta^{(3)}$ or higher.

All this does not mean however that we cannot embed  
the object having the dimension $D-3$, $D-4$, etc. into a D-dimensional
manifold -
simply in that case the object ``sees'' (at most)  two dimensions
whereas others form a product space {\it a la} Kaluza-Klein.
For example, a point particle (zero-dimensional object) can be embedded
into a four-dimensional spacetime but the Schwarzschild
metric is a product space
of $S^2$ and two-dimensional time-radius part hence one can say that a
point particle ``sees'' two dimensions.

Further, the case $D_E=2$ deserves for special treatment because
it is a limit case besides it comprises two-dimensional strings
in four-spacetime or four-vertices in six-dimensional spacetime
(the latters also were used in the early brane-world proposals 
\cite{akam}).
If $D_E=2$ then the surface curvature 2-form in eq. (\ref{eCurvDecomp})
is some 0-form times the 2-form $\drm\eta^{(1)}\wedge \drm\eta^{(2)}$.
Further, if we want to obtain this from somehow decomposed vielbein
the problem is how to find this ``somehow''.
The case $D_E=1$ is sharply distinct because there $\Sigma$ is
(D-1)-dimensional and hence one is able to introduce the notions
``on one side of $\Sigma$'' and ``on another side of $\Sigma$'' but
they are meaningless when $D_E > 1$ so we cannot start with
something like eq. (\ref{eSplitOm}).

So, let assume that $D_E=2$ and the embedding of (D-2)-dimensional
$\Sigma$ is described by two equations
\[
\eta^{(a)} (x) =0, \ a=1, 2 ,
\]
i.e., $\Sigma$ is the intersection of the two (D-1)-dimensional surfaces,
${\cal B}_1$ and ${\cal B}_2$,
described by the equations $\eta^{(1)} (x) =0$ and $\eta^{(2)} (x) =0$
respectively.
Instead of eq. (\ref{eSplitOm}) we assume
\be
\omega^\alpha = \omega^\alpha_{++} \theta_1 \theta_2  
+ \omega^\alpha_{+-} \theta_1 \theta_{-2}
+ \omega^\alpha_{-+} \theta_{-1} \theta_{2}
+ \omega^\alpha_{--} \theta_{-1} \theta_{-2},
\ee
where it is denoted $\theta_{\pm a} \equiv \theta (\pm \eta^{(a)})$ for
brevity,
in hope that later on we will find the restrictions for these $\omega$'s
because for unrestricted $\omega$'s this equation describes the two 
above-mentioned intersecting surfaces whereas we are 
interested in their intersection region ($\Sigma$) only. 

After taking external derivative we obtain
\ba
&&\drm \omega^\alpha =
\drm \omega^\alpha_{++} \theta_{1} \theta_{2} +
\drm \omega^\alpha_{+-} \theta_{1} \theta_{-2} +
\drm \omega^\alpha_{-+} \theta_{-1} \theta_{2} +
\drm \omega^\alpha_{--} \theta_{-1} \theta_{-2} + \nn\\
&&\qquad \quad 
(\Delta^\alpha_2 \theta_2 + \tilde\Delta^\alpha_2 \theta_{-2})
\wedge \drm \theta_1 +
(\Delta^\alpha_1 \theta_1 + \tilde\Delta^\alpha_1 \theta_{-1})
\wedge \drm \theta_2,
\ea
where
\ba
&&
\Delta_1 \equiv 
\left( \omega_{++} - \omega_{+-} \right)|_{{\cal B}_2}, \
\tilde\Delta_1 \equiv 
\left( \omega_{-+} - \omega_{--} \right)|_{{\cal B}_2}, \nn\\
&&
\Delta_2 \equiv 
\left( \omega_{++} - \omega_{-+} \right)|_{{\cal B}_1}, \
\tilde\Delta_2 \equiv 
\left( \omega_{+-} - \omega_{--} \right)|_{{\cal B}_1}, 
\ea
and, of course, it should not be forgotten that 
$\drm \theta_a = \delta (\eta^{(a)}) \drm \eta^{(a)}$.
Then after tedious but straightforward calculation we obtain that
the analogue of eq. (\ref{eGamDecomp}) is
\ba
&&\Gamma^\alpha_\beta =
\Gamma^{++ \alpha}_{\quad \beta} \theta_{1} \theta_{2} +
\Gamma^{+- \alpha}_{\quad \beta} \theta_{1} \theta_{-2} +
\Gamma^{-+ \alpha}_{\quad \beta} \theta_{-1} \theta_{2} +
\Gamma^{-- \alpha}_{\quad \beta} \theta_{-1} \theta_{-2} +  \nn\\
&&\qquad
(\Gamma_{(2)\beta}^{\quad \alpha} \theta_{2} + 
\tilde\Gamma_{(2)\beta}^{\quad \alpha} \theta_{-2}) \, \drm \theta_1
+
(\Gamma_{(1)\beta}^{\quad \alpha} \theta_{1} + 
\tilde\Gamma_{(1)\beta}^{\quad \alpha} \theta_{-1}) \, \drm \theta_2 , 
\lb{eGamDecomp2}
\ea
where $\Gamma^{\pm \pm \alpha}_{\quad \beta}$ are the standard connection
1-forms calculated from the corresponding $\omega_{\pm\pm}$, and
the numbered 0-forms $\Gamma^{\quad \ \, \alpha}_{(1,2)\beta}$ 
are the solutions of
the following linear equations
\ba
&& 
\theta (0) \sigma_a^\beta \Gamma_{(a) \beta}^{\quad \alpha}  
= \Delta_a^\alpha, \ \
\theta (0) \tilde \sigma_a^\beta \tilde\Gamma_{(a)\beta}^{\quad \alpha}  
= \tilde\Delta_a^\alpha,   \quad (a=1,2),
\ea
where it has been defined
\ba
&&
\sigma_1 \equiv 
\left( \omega_{++} + \omega_{+-} \right)|_{{\cal B}_2}, \
\tilde\sigma_1 \equiv 
\left( \omega_{-+} + \omega_{--} \right)|_{{\cal B}_2}, \nn\\
&&
\sigma_2 \equiv 
\left( \omega_{++} + \omega_{-+} \right)|_{{\cal B}_1}, \
\tilde\sigma_2 \equiv 
\left( \omega_{+-} + \omega_{--} \right)|_{{\cal B}_1}, 
\ea
besides we have not specified the value of the Heaviside 0-form when
its argument is zero.

To find the decomposed curvature 2-form we have to  take again the
external derivative of $\Gamma^\alpha_\beta$ and eventually we
obtain that the curvature form consists of the following
three parts -
the bulk part which does not contain delta-singularities,
the first brane part which describes the hypersurface
${\cal B}_1$ and is proportional to $\drm \theta_1$,
the second brane part which describes the hypersurface
${\cal B}_2$ and is proportional to $\drm \theta_2$,
and the 
intersection part 
which describes $\Sigma= {\cal B}_1 \cap {\cal B}_2$ and 
is a certain 0-form times the two-dimensional delta-``function''
$\drm \theta_1 \wedge \drm \theta_2$:
\ba
&&R^\alpha_\beta =
R^{++ \alpha}_{\quad \beta} \theta_{1} \theta_{2} +
R^{+- \alpha}_{\quad \beta} \theta_{1} \theta_{-2} +
R^{-+ \alpha}_{\quad \beta} \theta_{-1} \theta_{2} +
R^{-- \alpha}_{\quad \beta} \theta_{-1} \theta_{-2} +  \nn\\
&&\qquad \ \
\drm \theta_1 \wedge \,^{(1)}\!B_{\beta}^{\alpha}+
\drm \theta_2 \wedge \,^{(2)}\!B_{\beta}^{\alpha}+
\drm \theta_1 \wedge \drm \theta_2 \, S_{\beta}^{\alpha},
\lb{eCurvDecomp2}
\ea
where
\ba
^{(1)}\!B_{\beta}^{\alpha} &=&
\biggl[
       - \theta_{2} 
       (
        \Gamma^{++ \alpha}_{\quad \beta} 
        - \Gamma^{- + \alpha}_{\quad \beta}
        + \drm \Gamma_{(2)\beta}^{\quad \alpha}
       )
       - \theta_{-2} 
       (
        \Gamma^{+- \alpha}_{\quad \beta} 
        - \Gamma^{-- \alpha}_{\quad \beta}
        + \drm \tilde\Gamma_{(2)\beta}^{\quad \alpha}
       ) 
       +
\nn\\
&& \
       \theta_{2} \theta (0)
       (
        \Gamma^{++ \gamma}_{\quad \beta} 
        + \Gamma^{-+ \gamma}_{\quad \beta}
       ) \Gamma^{\quad \alpha}_{(2)\gamma} 
       +
       \theta_{-2} \theta (0)
       (
        \Gamma^{+- \gamma}_{\quad \beta} 
        + \Gamma^{-- \gamma}_{\quad \beta}
       ) \tilde\Gamma^{\quad \alpha}_{(2)\gamma}
       -
\nn\\
&& \
       \theta_{2} \theta (0)
       (
        \Gamma^{++ \alpha}_{\quad \gamma} 
        + \Gamma^{-+ \alpha}_{\quad \gamma}
       ) \Gamma^{\quad \gamma}_{(2)\beta} 
       -
       \theta_{-2} \theta (0)
       (
        \Gamma^{+- \alpha}_{\quad \gamma} 
        + \Gamma^{-- \alpha}_{\quad \gamma}
       ) \tilde\Gamma^{\quad \gamma}_{(2)\beta}
\biggr]_{{\cal B}_1},
\ea
\ba
^{(2)}\!B^{\alpha}_{\beta} &=&
\biggl[
       - \theta_{1} 
       (
        \Gamma^{++ \alpha}_{\quad \beta} 
        - \Gamma^{+- \alpha}_{\quad \beta}
        + \drm \Gamma_{(1)\beta}^{\quad \alpha}
       )
       - \theta_{-1} 
       (
        \Gamma^{-+ \alpha}_{\quad \beta} 
        - \Gamma^{-- \alpha}_{\quad \beta}
        + \drm \tilde\Gamma_{(1)\beta}^{\quad \alpha}
       ) 
       +
\nn\\
&& \
       \theta_{1} \theta (0)
       (
        \Gamma^{++ \gamma}_{\quad \beta} 
        + \Gamma^{+- \gamma}_{\quad \beta}
       ) \Gamma^{\quad \alpha}_{(1)\gamma} 
       +
       \theta_{-1} \theta (0)
       (
        \Gamma^{-+ \gamma}_{\quad \beta} 
        + \Gamma^{-- \gamma}_{\quad \beta}
       ) \tilde\Gamma^{\quad \alpha}_{(1)\gamma}
       -
\nn\\
&& \
       \theta_{1} \theta (0)
       (
        \Gamma^{++ \alpha}_{\quad \gamma} 
        + \Gamma^{+- \alpha}_{\quad \gamma}
       ) \Gamma^{\quad \gamma}_{(1)\beta} 
       -
       \theta_{-1} \theta (0)
       (
        \Gamma^{-+ \alpha}_{\quad \gamma} 
        + \Gamma^{-- \alpha}_{\quad \gamma}
       ) \tilde\Gamma^{\quad \gamma}_{(1)\beta}
\biggr]_{{\cal B}_2},
\ea
and
\ba
S^\alpha_\beta &=&
\biggl[
       \Gamma^{\quad \alpha}_{(1)\beta} -
       \Gamma^{\quad \alpha}_{(2)\beta}
       - 
       \left(
       \tilde\Gamma^{\quad \alpha}_{(1)\beta} -
       \tilde\Gamma^{\quad \alpha}_{(2)\beta} 
       \right) +
\nn\\&&
      \theta(0)^2 
      \left(
            \Gamma^{\quad \alpha}_{(2)\gamma}
            \Gamma^{\quad \gamma}_{(1)\beta}
            -
            \Gamma^{\quad \alpha}_{(1)\gamma}
            \Gamma^{\quad \gamma}_{(2)\beta}
      \right)
      +
      \theta(0)^2 
      \left(
            \Gamma^{\quad \alpha}_{(2)\gamma}
            \tilde\Gamma^{\quad \gamma}_{(1)\beta}
            -
            \tilde\Gamma^{\quad \alpha}_{(1)\gamma}
            \Gamma^{\quad \gamma}_{(2)\beta}
      \right) +
\nn\\&&
      \theta(0)^2 
      \left(
            \tilde\Gamma^{\quad \alpha}_{(2)\gamma}
            \Gamma^{\quad \gamma}_{(1)\beta}
            -
            \Gamma^{\quad \alpha}_{(1)\gamma}
            \tilde\Gamma^{\quad \gamma}_{(2)\beta}
      \right)
      +
      \theta(0)^2 
      \left(
            \tilde\Gamma^{\quad \alpha}_{(2)\gamma}
            \tilde\Gamma^{\quad \gamma}_{(1)\beta}
            -
            \tilde\Gamma^{\quad \alpha}_{(1)\gamma}
            \tilde\Gamma^{\quad \gamma}_{(2)\beta}
      \right) 
\biggr]_{\Sigma = {\cal B}_1\cap {\cal B}_2}.
\ea
Further, if one wishes that eq. (\ref{eCurvDecomp2}) describes only
the intersection $\Sigma$ one must impose on the brane 1-forms
the following two additional restrictions:
\be
^{(a)}\!B^{\alpha}_{\beta} = \,^{(a)}\!b^{\alpha}_{\beta}\, \drm\eta^{(a)},
\ \
(a=1,2),
\ee
with $\,^{(a)}\!b^{\alpha}_{\beta}$ being the arbitrary functions
containing $\theta$'s.

\section{Discussion}\lb{s-dis}

Let us discuss now the features of singular ($V+E$)-submanifolds in details.
Some of these features are drastically new with respect to
both the physics of Kaluza-Klein type theories and that of the 
($D-1$)-dimensional singular embeddings including ($D-2$)-branes
as a special case.
Then the two points of view on them, optimistic (constructive) 
and pessimistic (destructive), will be outlined.
So, the features are:

(i) {\it The distinction of physics of singular manifolds from that
of Kaluza-Klein dimensional reduction}.\\
It is well-known that nowadays 
the dimension of our visible Universe is regarded
to be four  due to many reasons,
so higher-dimensional theories are obliged to eventually
hide extra spacetime dimensions to fit the experimental data.
For instance, the main style of thinking in the KK type models 
is to consider
the {\it smooth} everywhere (except perhaps a finite number
of points) high-$D$ spacetime, decompose its values into the
$E$ 
and $V$ (yet non-singular) parts, associating the
former one with internal degrees of
freedom, and then assume extra dimension compact
with small size.
Unlike this the singular manifold requires neither compactness nor 
hugeness of $E$-coordinates.
It is entity which lives (moves and warps)
inside a parent manifold (the latter becomes
to be only $C^0$ in the vicinity of $V$), and 
has very own intrinsic geometry,
matter, Standard Model (which is confined on $V$ unlike gravitation), etc.
However, the parent high-$D$ manifold affects
both the internal physics and external motion of the baby manifold,
so as a consequence of this, there appears a number
of distinctive features which are
very specific for the singular submanifolds.

The higher-dimensional generation of internal degrees of freedom 
in the Universe $V$ (in addition to the fibers of group spaces over
the tangent bundle $T_V$) exists also in the KK type
theories \cite{vla} as the phenomenon
induced by $E$-metric $g_{a b}$ and it is independent of whether we 
have the singular split $\Sigma = V \cup E$ or only 
the decomposition
$T_\Sigma=T_V \oplus T_E$.
Therefore, KK mechanism also works on a singular submanifold.
However, the physics of latter is determined not only by confined 
matter (including the SM with its fiber bundles), by
dynamics of $V$ as a whole 
(also specific for singular submanifolds only), and by 
projected bulk $\Sigma$-gravity,
but also by the high-$D$ boundary effects including the anisotropy
caused by the presence of multiple normals.

One may feel some vague analogy of this effect with
the holography principle \cite{tho,wit} which is also a boundary effect.
However, at the present stage of the theory this connection
yet seems to be too dim because the holography principle in its most
radical form suggests that the information about volume processes
is stored on the surface
whereas the multi-normal embedding approach in initial form means
the high-$D$ mechanism of generation of
internal degrees of freedom without introducing the fiber of
internal symmetry groups.

(ii) {\it The (weak) violation of $V$-relativistic covariance 
and restricted structure of the parent manifold}.\\
The thorough look at the induced-gravity equations above and 
feature (i) reveals that the relativistic covariance  is violated
on the baby manifold $V$ while preserved in the parent spacetime $\Sigma$.
It can easily be seen that the effective stress-energy tensor
will contain the terms which are not
$V$-tensors.
Generally speaking, the violation of relativity takes
place for KK theories as well 
(because of the $(V+E)$-decomposition formulae 
are the same for both singular and non-singular submanifolds),
but in that case the $(V+E)$-decomposition is at most than 
the mathematical relabeling of a high-$D$
manifold, and hence the
equations have no physical (induced-gravity) sense there.
The relativity violation takes place also when considering 
$(D-1)$-embeddings  ($(D-2)$-branes) both in the junction and 
distributional approaches because it assumes the 
implicit separation
of the extra dimension which is assumed to be orthogonal, i.e.,
similar to the gauss/synchronous coordinate.
In the consistent theory of $(D-1)$-embeddings 
it is impossible
to introduce the crucial definition of external curvature 
without this orthogonality. 
The assumption of orthogonality also means that if $D>4$
we restrict ourselves to the
spaces of a special ``block-orthogonal'' type:
while the standard ($3+1$) decomposition can be justified as the
appropriately chosen coordinate system and hence implies no
restrictions on the whole 4D manifold,
we cannot say the same about 5D, 6D manifolds.

(iii) {\it The reformulation of the gravitational energy-momentum 
tensor problem}.\\
The embedded-world viewpoint can be applied to the 
old-standing problem of the conserved
energy-momentum tensor for gravitational field (CGEMT) which is
sometimes regarded as the main disadvantage of general relativity.
Indeed, once we have imagined our Universe as 
the singular embedding inside the parent Meta-Universe
there is no physical sense to require the
conservation of the four-dimensional GEMT because the 
four-Universe explicitly becomes a  gravitationally 
non-conservative system:
gravity is not confined inside the four-manifold.
The CGEMT problem is thus reduced to that of the higher-dimensional CGEMT.
However, we can adjust the internal geometry and external dynamics
of $V$ in such a way that the high-$D$ spacetime 
becomes flat or, at least, of constant curvature. 
Then the GEMT problem vanishes as well for the $D$-spacetime.
The question is thus whether it is possible to do so that the 
perturbations
of high-$D$ metric caused by the matter on $V$ and the
external dynamics of $V$ as a whole cancel each other out.
Considering the emergent huge freedom of the as-a-whole 
external dynamics of singular submanifolds
it seems to be possible, moreover, in a non-unique way.

(iv) {\it The parent manifold is not the manifold}.\\
The rigorous definition of the (differentiable) manifold require smoothness
(including the smooth sewing of all the parts)
and hence local diffeomorphicity to ${\bf R}^D$.
However, in the vicinity of a singular submanifold the smoothness of
the parent ``manifold'' 
breaks down \cite{zlo004}, 
major definitions fail, and therefore still there is a 
question what is the physics ``on the edge'' \cite{bv}.\\

After we have  enumerated all the main objective peculiarities of
the singular manifolds it is time to represent the subjective
points of view on some of them.

$\bullet $ 
The {\it optimistic} (constructive) viewpoint suggests the following.
The feature (i) does mean that the singular submanifold theories
and brane-world models
is new and promising mathematical tool and model of our Universe
as part of the Meta-Universe.
It provides us with opportunities to study the physical embeddings
in the high-dimensional (even infinitely-dimensional) spaces.
The feature (ii) is not the problem because the relativity
holds for the whole spacetime $\Sigma$ whereas the contributions violating
the $V$-covariance can be regarded to
give only the higher-order corrections to the induced
Einstein equations ruling over the $V$(isible) Universe.

$\bullet $ 
If the hopes set upon the optimistic viewpoint 
will not be justified we should recall some disadvantages of
the brane-world paradigm and seriously consider 
the {\it pessimistic} point of view.
Indeed, 
looking back in time and comparing this paradigm to that of the 
Kaluza-Klein compactification we can see a number of defects.
Apart from the problems mentioned above 
the serious one is that the non-uniqueness, which was inherent to the 
KK theories, is even more amplified due to the appearance of
the external dynamics of the baby manifold as a whole.\footnote{Besides, 
we cannot completely replace the
KK mechanism by the brane-world one because
following the aforesaid the four-Universe can be singularly embedded 
only in five- and six-dimensional space so if we want to
consider its embedding in 10D then other dimensions must be included but 
as a product space only.}
The researcher modeling physical reality by means of brane
embeddings can obtain everything he wants and in several ways,
and it does not seem to be a good sign because this
decreases the foretelling ability of the theory.
On the other hand, 
the experimental detectability of the high-$D$ phenomena projected
onto our Universe is a separate large problem \cite{aadd,pes}.

\section*{Acknowledgments}
I would like to thank Edward Teo 
(DAMTP, Univ. of Cambridge and Natl. Univ. of Singapore)
and Cristian Stelea (Univ. ``Alexandru Ioan Cuza'', Iasi, Romania)
for suggesting the theme
and helpful discussions.

\def\AnP{Ann. Phys.}
\def\APP{Acta Phys. Polon.}
\def\CJP{Czech. J. Phys.}
\def\CMPh{Commun. Math. Phys.}
\def\CQG {Class. Quantum Grav.}
\def\EPL  {Europhys. Lett.}
\def\IJMP  {Int. J. Mod. Phys.}
\def\JMP{J. Math. Phys.}
\def\JPh{J. Phys.}
\def\FP{Fortschr. Phys.}
\def\GRG {Gen. Relativ. Gravit.}
\def\GC {Gravit. Cosmol.}
\def\LMPh {Lett. Math. Phys.}
\def\MPL  {Mod. Phys. Lett.}
\def\NPh  {Nucl. Phys.}
\def\PhE  {Phys.Essays}
\def\PhL  {Phys. Lett.}
\def\PhR  {Phys. Rev.}
\def\PhRL {Phys. Rev. Lett.}
\def\PhRp {Phys. Rep.}
\def\PTP {Prog. Theor. Phys.}
\def\NCim {Nuovo Cimento}
\def\TMF {Teor. Mat. Fiz.}
\def\prp {report}
\def\Prp {Report}

%\def\jn#1#2#3#4#5{{#1}{#2} {\bf #3}, {#4} {(#5)}} %PRD
%\def\jn#1#2#3#4#5{{#1}{#2} {#3} {(#5)} {#4}}   %PLB style
% #1 tittle  #2 ser  #3 vol  #4 page  #5 year

%\def\boo#1#2#3#4#5{{\it #1} ({#2}, {#3}, {#4}){#5}}
%\def\boo#1#2#3#4#5{ #1 ({#2}, {#3}, {#4}){#5}}  %PLB style
% #1 tittle  #2 publisher  #3 place  #4 year  #5 page/, p.789/

%\def\jn#1#2#3#4#5{{#1}{#2} {\bf #3}, {#4} {(#5)}}
% #1 tittle  #2 ser  #3 vol  #4 page  #5 year
%\def\boo#1#2#3#4#5{{\it #1} ({#2}, {#3}, {#4}){#5}}
% #1 tittle  #2 publisher  #3 place  #4 year  #5 page/, p.789/

%IOP
\def\jn#1#2#3#4#5{#5 {\it #1}{#2} {\bf #3} {#4}}
% #1 tittle  #2 ser  #3 vol  #4 page  #5 year
\def\boo#1#2#3#4#5{#4 {\it #1} ({#3}: {#2}){#5}}
% #1 tittle  #2 publisher  #3 place  #4 year  #5 page/, p 789/

\end{document}